\documentclass[fleqn,usenatbib,useAMS]{mnras}

\usepackage{graphicx}	
\usepackage{amsmath}	
\usepackage{amssymb}	
\usepackage{multicol}        
\usepackage{bm}		
\usepackage{pdflscape}	
\usepackage[normalem]{ulem}
\usepackage{tikz}
\usetikzlibrary{shapes.geometric, arrows}

\tikzstyle{input} = [rectangle, minimum width=1cm, minimum height=1cm, text centered, draw=black, fill=blue!10]

\tikzstyle{output} = [rectangle, minimum width=1cm, minimum height=1cm, text centered, draw=black, fill=white]

\tikzstyle{process} = [rectangle, minimum width=1cm, minimum height=1cm, text centered, draw=black, fill=orange!30]

\tikzstyle{decision} = [rectangle, minimum width=1cm, minimum height=1cm, text centered, draw=black, fill=green!30]

\tikzstyle{arrow} = [thick,->,>=stealth]

\usepackage[T1]{fontenc}
\usepackage{ae,aecompl}

\usepackage{newtxtext,newtxmath}

\title[\texttt{matryoshka} II]{\texttt{matryoshka} II: Accelerating Effective Field Theory Analyses of the Galaxy Power Spectrum}

\author[J. Donald-McCann et al.]{
Jamie Donald-McCann,$^{1}$\thanks{E-mail: jamie.donald-mccann@port.ac.uk}
Kazuya Koyama,$^{1}$
Florian Beutler$^{2}$
\\
$^{1}$Institute of Cosmology \& Gravitation, University of Portsmouth, Dennis Sciama Building, Portsmouth, PO1 3FX, UK\\
$^{2}$Institute for Astronomy, University of Edinburgh, Royal Observatory, Blackford Hill, Edinburgh EH9 3HJ, UK
}

\date{Accepted XXX. Received YYY; in original form ZZZ}

\pubyear{2022}

\begin{document}
\label{firstpage}
\pagerange{\pageref{firstpage}--\pageref{lastpage}}
\maketitle

\begin{abstract}
In this paper we present an extension to the \texttt{matryoshka} suite of neural-network-based emulators. The new editions have been developed to accelerate EFTofLSS analyses of galaxy power spectrum multipoles in redshift space. They are collectively referred to as the \texttt{EFTEMU}. We test the \texttt{EFTEMU} at the power spectrum level and achieve a prediction accuracy of better than 1\% with BOSS-like bias parameters and counterterms on scales $0.001\ h\ \mathrm{Mpc}^{-1} \leq k \leq 0.19\ h\ \mathrm{Mpc}^{-1}$. We also run a series of mock full shape analyses to test the performance of the \texttt{EFTEMU} when carrying out parameter inference. Through these mock analyses we verify that the \texttt{EFTEMU} recovers the true cosmology within $1\sigma$ at several redshifts ($z=[0.38,0.51,0.61]$), and with several noise levels (the most stringent of which is Gaussian covariance associated with a volume of $5000^3 \ \mathrm{Mpc}^3 \ h^{-3}$). We compare the mock inference results from the \texttt{EFTEMU} to those obtained with a fully analytic EFTofLSS model and again find no significant bias, whilst speeding up the inference by three orders of magnitude. The \texttt{EFTEMU} is publicly available as part of the \texttt{matryoshka} \texttt{Python} package.
\end{abstract}





\section{Introduction}
The use of emulators is becoming commonplace in many forms of cosmological analysis \citep{chapman_completed_2021, kobayashi_full-shape_2021, zurcher_dark_2021, white_cosmological_2021, miyatake_cosmological_2021}. These emulators can be thought of as sophisticated interpolation schemes that aim to approximate a computationally expensive model given a set of example outputs: providing fast predictions whilst maintaining accuracy.

Developing emulators to facilitate cosmological analyses was initially proposed as a method to overcome the huge computational cost of numerical simulations \citep{heitmann_cosmic_2006}. Numerical simulations provide predictions for clustering that are accurate on small nonlinear scales \citep{schneider_matter_2016, vogelsberger_cosmological_2019, angulo_large-scale_2021}\textcolor{teal}{;} however as most cosmological analyses will require many simulations their cost prohibits their use. There has been a lot of work focused on development of emulators in this context in recent years, and emulators have been produced for many statistics, including: the matter power spectrum \citep{heitmann_coyote_2009, agarwal_pkann_2014, giblin_road_2019, euclid_collaboration_euclid_2019, angulo_bacco_2020}, the halo power spectrum in redshift space \citep{kobayashi_accurate_2020}, the galaxy power spectrum \citep{kwan_cosmic_2015}, the halo correlation function \citep{nishimichi_dark_2019}, and the galaxy correlation function in redshift space \citep{zhai_aemulus_2019}.

Analytical predictions of the large-scale structure are more computationally efficient than those from numerical simulations. State of the art perturbation theory based models \citep{ivanov_cosmological_2020, philcox_combining_2020, damico_cosmological_2020, chen_consistent_2020} are able to produce predictions for the power spectrum multipoles that are accurate on quasi-nonlinear scales in a few seconds. The speed of these models makes it feasible to use them directly when conducting cosmological inference. However a typical Markov chain Monte Carlo (MCMC) analysis using one of these perturbative models has $\mathcal{O}(10)$ free parameters, requiring $\mathcal{O}(10^{5}-10^{6})$ model evaluations to converge. As such these analyses still require considerable computational resources. In recent years many works have looked at using the idea of emulation to accelerate analytic predictions not just those coming from numerical simulations \citep{arico_accelerating_2021, derose_neural_2021, mancini_cosmopower_2022}. Emulation of  analytic predictions can greatly reduce the computational cost for parameter inference. This can allow for different analysis setups to be easily explored (i.e. prior choice), and removes the need for computing clusters to do inference. 

In \citet{donald-mccann_textttmatryoshka_2022} we introduced \texttt{matryoshka}, a suite of neural-network-based emulators. In this work we add a set of new emulators to \texttt{matryoshka}. These new emulators have been developed to accelerate effective field theory of large scale structure (EFTofLSS) predictions. We collectively refer to these new emulators as the \texttt{EFTEMU}. The structure of this paper is as follows: In section \ref{sec:training_data} we outline the input parameters of the \texttt{EFTEMU} and the data we generate for training. In section \ref{sec:neural_nets} we discuss the training procedure for the individual component emulators that make up the \texttt{EFTEMU}, along with the prediction accuracy at the power spectrum level. In section \ref{sec:mock_analyses} we present a series of mock analyses designed to test the accuracy of the \texttt{EFTEMU}. In section \ref{sec:performance} we quantify the computational performance that the \texttt{EFTEMU} provides. We conclude in section \ref{sec:conclusions}.

\section{Generating Training Data}
\label{sec:training_data}
The \texttt{EFTEMU} has been developed to approximate an EFTofLSS model for the galaxy power spectrum multipoles. \citet{damico_cosmological_2020} show that the galaxy power spectrum multipoles can be predicted by combining a series of bias parameters and counterterms (throughout the rest of this paper we refer to both when we mention `bias parameters') with components $P_{n,l}$ that solely depend on cosmology, such that 
\begin{equation}
    P_l(k) = \sum_n b_n P_{n,l}(k)\ .
    \label{eq:multipole_sum}
\end{equation}
Each $P_{n,l}$ in the equation above represents the result of a convolution of the linear matter power spectrum with a redshift space galaxy density or velocity kernel (or combination of results of convolutions with different kernels). Each $b_n$ represents a bias parameter (or combination of multiple bias parameters) for the corresponding $P_{n,l}$ component. For more details on the form of EFTofLSS model and the convolution kernels we refer the reader to \citet{damico_cosmological_2020} and the references therein.

Emulating these $P_{n,l}$ components rather than the galaxy multipoles themselves is advantageous for two reasons. Firstly it means that no prior is required on bias parameters when making predictions with the \texttt{EFTEMU}. Secondly and more importantly the training volume is significantly smaller when the bias parameters are not included as input parameters for the \texttt{EFTEMU}. This smaller training volume means we require less training data to reach a given level of accuracy. For this work, we use \texttt{PyBird} \citep{damico_limits_2021} to calculate all $P_{n,l}$ components and \texttt{CLASS} \citep{lesgourgues_cosmic_2011} to calculate the linear matter power spectrum. We choose \texttt{PyBird} because it is fast, documented, and allows for the $P_{n,l}$ terms to be easily extracted. We note that the use of this specific code is not mandatory. We could use any code that can calculate the $P_{n,l}$ components of equation \ref{eq:multipole_sum}.

\begin{figure}
	\includegraphics[width=\columnwidth]{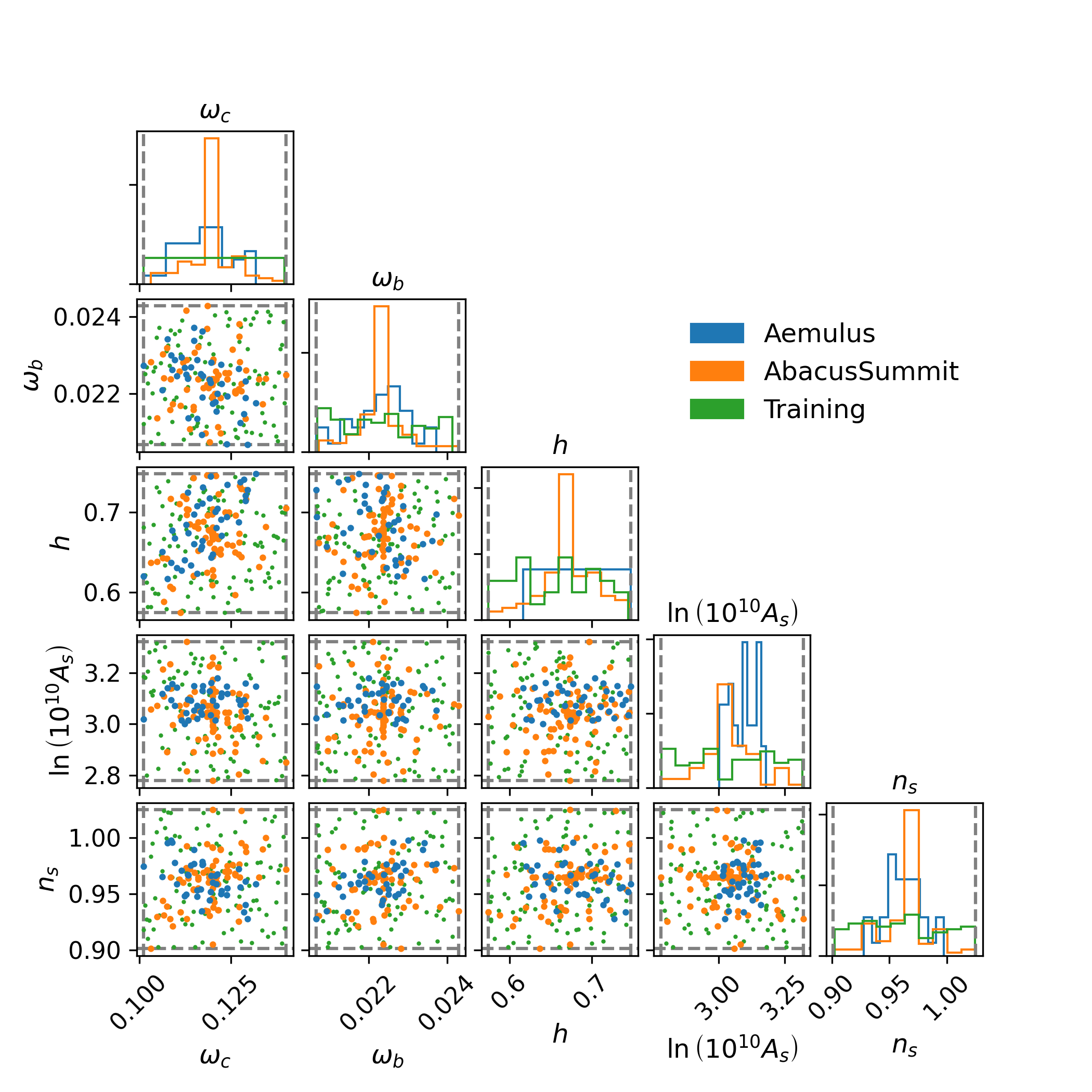}
    \caption{Training space for the \texttt{EFTEMU}. The blue and orange points show the locations of Aemulus and AbacusSummit simulation samples. These simulation samples are used to inform the training space for the \texttt{EFTEMU}, as described in section \ref{sec:training_data}. The green points show 100 random samples from the \texttt{EFTEMU} training space, and the grey dashed lines show the extremes. This covers an approximately 13--24$\sigma$ region around the Planck 2018 $\Lambda$CDM cosmology.}
    \label{fig:sample_space}
\end{figure}

The \texttt{EFTEMU} takes five $\Lambda$CDM cosmological parameters as input; the density of cold dark matter $\Omega_c h^2$, the density of baryonic matter $\Omega_b h^2$, the dimensionless Hubble parameter $h = H_0 \ / \ (100 \ \mathrm{km}\ \mathrm{s}^{-1}\ \mathrm{Mpc}^{-1})$, and the amplitude and tilt of the primordial power spectrum  $\ln{\left(10^{10}A_s\right)}$ and $n_s$. When generating samples from this $\Lambda$CDM parameter space we have to impose priors. For this work these priors are defined by public suites of numerical simulations, those being the Aemulus suite \citep{derose_aemulus_2019} and the AbacusSummit suite \citep{maksimova_abacussummit_2021}. We concatenate their shared $\Lambda$CDM parameters and generate 10,000 Latin-hypercube samples \citep{mckay_comparison_1979} in the region covered by the simulation samples. As both the Aemulus and AbacusSummit suites sample from beyond $\Lambda$CDM parameter spaces the ranges for the $\Lambda$CDM parameters are large. The training space covers an approximately 13--24$\sigma$ region around the most recent Planck $\Lambda$CDM TT, TE, EE + lowE + lensing + BAO results \citep[table 2 in][henceforth Planck 2018]{planck_collaboration_planck_2020}. Figure \ref{fig:sample_space} shows the Aemulus and AbacusSummit samples, along with 100 random samples from the \texttt{EFTEMU} training space.

We calculate the $P_{n,l}$ components for all 10,000 samples at three redshifts $z=[0.38,0.51,0.61]$, covering scales $0.001\ h\ \mathrm{Mpc}^{-1} \leq k \leq 0.19\ h\ \mathrm{Mpc}^{-1}$.\footnote{We calculate $P_{n,l}$ with \texttt{PyBird} up to $k=0.3\ h\ \mathrm{Mpc}^{-1}$ with \texttt{optiresum=False} (\texttt{True} focuses resummation on the BAO peak). However a small amount of numerical noise on scales $>0.19\ h\ \mathrm{Mpc}^{-1}$ detrimentally impacted training so we discarded these scales. Throughout this work we use \texttt{PyBird v0.1} on the \texttt{master} branch \url{https://github.com/pierrexyz/pybird}} Only 8,000 were used for training; the other 2,000 were used to test the prediction accuracy of the multipoles (see section \ref{subsec:per_err}).

\section{Training the Neural Networks}
\label{sec:neural_nets}
In this section we discuss the training procedure for the set of neural-network-based emulators that forms the \texttt{EFTEMU}. Each emulator in the set is a \textit{component emulator}, and takes the form of a simple fully connected neural network (NN). The output of all the component emulators is combined when making a prediction for the galaxy power spectrum multipoles. NNs learn a target function, in this case the bias-independent components of the multipole predictions $P_{n,l}$, by optimising a set of weights $\bm{w}$. It is common practice to preprocess the training data before optimising $\bm{w}$. For this work, the preprocessing involves rescaling all target functions and input variables such that they lie within the range $[0,1]$. This preprocessing ensures that all outputs contribute equally when optimising $\bm{w}$, and also improves training stability by keeping the magnitude of $\bm{w}$ small\footnote{High magnitude target functions can lead to large weights, and this can in turn lead to the gradients of the loss function with respect to these weights becoming large. Large gradients can be detrimental to training as they are used to inform the updates of the weights.}. We note that prior to this rescaling we remove any scales at which $P_{n,l}(k)=0$ for all samples in the training set.

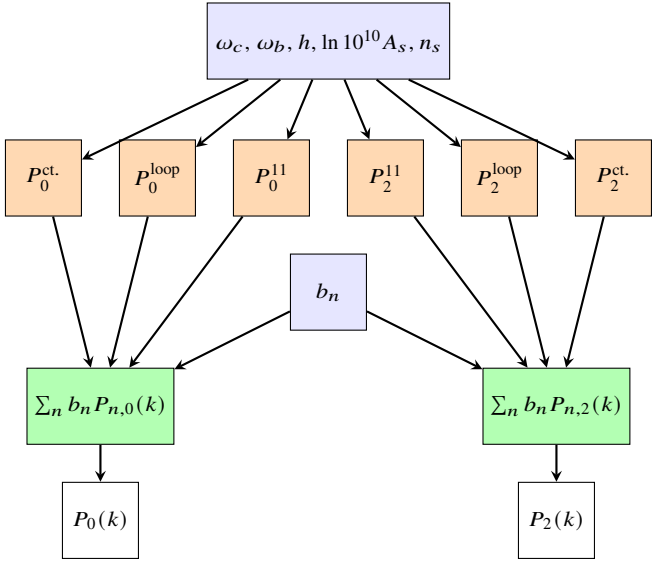
\begin{figure}

\begin{tikzpicture}[node distance=1.5cm]

\node (start) [input] {$\omega_c$, $\omega_b$, $h$, $\ln{10^{10}A_s}$, $n_s$};

\node (g01) [process, below of=start, xshift=-0.75cm, yshift=-0.3cm] {$P^{11}_0$};
\node (g02) [process, below of=start, xshift=-2.25cm, yshift=-0.3cm] {$P^\mathrm{loop}_0$};
\node (g03) [process, below of=start, xshift=-3.75cm, yshift=-0.3cm] {$P^\mathrm{ct.}_0$};

\node (g21) [process, below of=start, xshift=0.75cm, yshift=-0.3cm] {$P^{11}_2$};
\node (g22) [process, below of=start, xshift=2.25cm, yshift=-0.3cm] {$P^\mathrm{loop}_2$};
\node (g23) [process, below of=start, xshift=3.75cm, yshift=-0.3cm] {$P^\mathrm{ct.}_2$};

\node (biasin) [input, below of=g01, xshift=0.75cm] {$b_n$};

\node (add0) [decision, below of=biasin, xshift=-3cm]
{$\sum_n b_n P_{n,0}(k)$};
\node (add2) [decision, below of=biasin, xshift=3cm]
{$\sum_n b_n P_{n,2}(k)$};

\node (P0out) [output, below of=add0] {$P_0(k)$};
\node (P2out) [output, below of=add2] {$P_2(k)$};

\draw [arrow] (start) -- (g01);
\draw [arrow] (start) -- (g02);
\draw [arrow] (start) -- (g03);
\draw [arrow] (start) -- (g21);
\draw [arrow] (start) -- (g22);
\draw [arrow] (start) -- (g23);

\draw [arrow] (g21) -- (add2);
\draw [arrow] (g22) -- (add2);
\draw [arrow] (g23) -- (add2);

\draw [arrow] (g01) -- (add0);
\draw [arrow] (g02) -- (add0);
\draw [arrow] (g03) -- (add0);

\draw [arrow] (biasin) -- (add0);
\draw [arrow] (biasin) -- (add2);

\draw [arrow] (add0) -- (P0out);
\draw [arrow] (add2) -- (P2out);

\end{tikzpicture}

\caption{Flowchart visualising how each of the component emulators is used together when making predictions for the monopole $P_0(k)$ and quadrupole $P_2(k)$ of the galaxy power spectrum. Pale blue boxes represent inputs, orange boxes component emulators, green boxes analytic operations, and white boxes outputs. We can see that there are three component emulators per multipole. Each component emulator produces predictions for a group of bias-independent $P_{n,l}$ components (for more details about the grouping see section \ref{sec:neural_nets}). The predictions from the component emulators are then combined with the bias parameters analytically to produce a prediction for the multipoles.}
\label{fig:flowchart}
\end{figure}

Rather than training individual component emulators for each $P_{n,l}$ we decide to train component emulators to make predictions for groups of $P_{n,l}$. There are 21 $P_{n,l}$ for both the monopole and the quadrupole; these 21 $P_{n,l}$ are split into three groups for each multipole. The first group contains linear contributions to the multipoles that are dominant on large scales and will be combined with the linear bias $b_1$ (this group is referred to as $P^{11}_l$). The second contains non-linear \textit{loop} contributions that become more important on small scales and will be combined with higher order bias parameters (this group is referred to as $P^\mathrm{loop}_l$). The third contains counterterm contributions which will be combined with at least one counterterm, and like the loop contributions will become more significant on small scales (this group is referred to as $P^\mathrm{ct.}_l$). The $P_{n,l}$ in each group are combined into a single vector that forms the target function for the relevant component emulator. When predictions are made with the \texttt{EFTEMU} the output of the component emulators is split into the relevant $P_{n,l}$ so they can be combined with the bias parameters. Figure \ref{fig:flowchart} shows a flowchart visualising how the six component emulators are used together when making predictions. Emulating groups in this way aids in prediction speed as a smaller number of NNs need to be evaluated when producing predictions. It also reduces the memory usage of the \texttt{EFTEMU} as a smaller number of weights need to be loaded into memory.

All the NNs are built with \texttt{TensorFlow} \citep{tensorflow_developers_tensorflow_2021}. All the NNs are shallow, each only having two fully-connected hidden layers. Both hidden layers have \texttt{ReLU} \citep{agarap_deep_2019} activation functions. The number of nodes in each hidden layer depends on the component emulator. For those that are emulating groups $P^{11}_l$ and $P^\mathrm{ct.}_l$ each hidden layer has 200 nodes. For those emulating the $P^\mathrm{loop}_l$ group each hidden layer has 400 nodes. A larger number of nodes in the hidden layers is required as the functional form of the $P_{n,l}$ in $P^\mathrm{loop}_l$ is more complex than those in $P^{11}_l$ and $P^\mathrm{ct.}_l$. We train the NNs for a maximum of 10,000 epochs with a batch size of 100. One epoch represents a complete pass through the training set. The batch size is the number of training samples used for each update of the NN weights. The weights are optimised using an \texttt{Adam} \citep{kingma_adam:_2017} optimiser. This optimisation involves comparing predictions made by the NNs to the preprocessed true $P_{n,l}$ components via a \textit{loss function}. For this work, we use a mean squared error loss function. The \textit{learning rate} controls how much the weights are changed on each update; we start training with a learning rate of 0.013. If there is a plateau in the loss function for more than ten epochs we reduce the learning rate by a factor of 0.1, unless the learning rate has already reached a value of 0.0001\footnote{Although \texttt{Adam} is already an adaptive optimiser we find that models reach a lower training loss when reducing the learning rate in this way.}. If there is a plateau in the loss function for more than 20 epochs we terminate training early. 

The number of hidden layers and nodes, the batch size, and the learning rate all represent hyper-parameters of the NNs that need to be tuned. For this work the hyper-parameters were adjusted manually, and the impact on the loss function observed. The set of hyper-parameters that resulted in the minimum loss was selected. There are more sophisticated methods for hyper-parameter tuning, such as Bayesian optimisation \citep{snoek_practical_2012}. However, they were not required to achieve predictions for $P_0$ and $P_2$ that agree with \texttt{PyBird} within 1\% for 68\% of test cases (i.e. 1\% at the 1$\sigma$ level) with arbitrary bias parameters.

\subsection{Power-spectrum-level prediction accuracy}
\label{subsec:per_err}

We test the prediction accuracy of the multipoles using the unseen test set. We use three sets of bias parameters ($b_n$ in equation \ref{eq:multipole_sum}) for these tests. The first set is a random draw from the prior defined in equation 3.18 of \citet{damico_cosmological_2020}. This random draw is different for each of the 2000 samples in the test set. The second and third sets of bias parameters are the best fit LOWZ NGC and CMASS NGC from \citet{damico_cosmological_2020} respectively. Testing the accuracy with the random bias parameters gives an idea of the prediction accuracy over the entire prior volume. Using the best fit parameters gives us an idea for the prediction accuracy for a more realistic set of bias parameters as a lot of the random combinations return multipoles that are nothing like what have been previously observed ($P_0$ going negative for example).

\begin{figure*}
	\includegraphics[width=\linewidth]{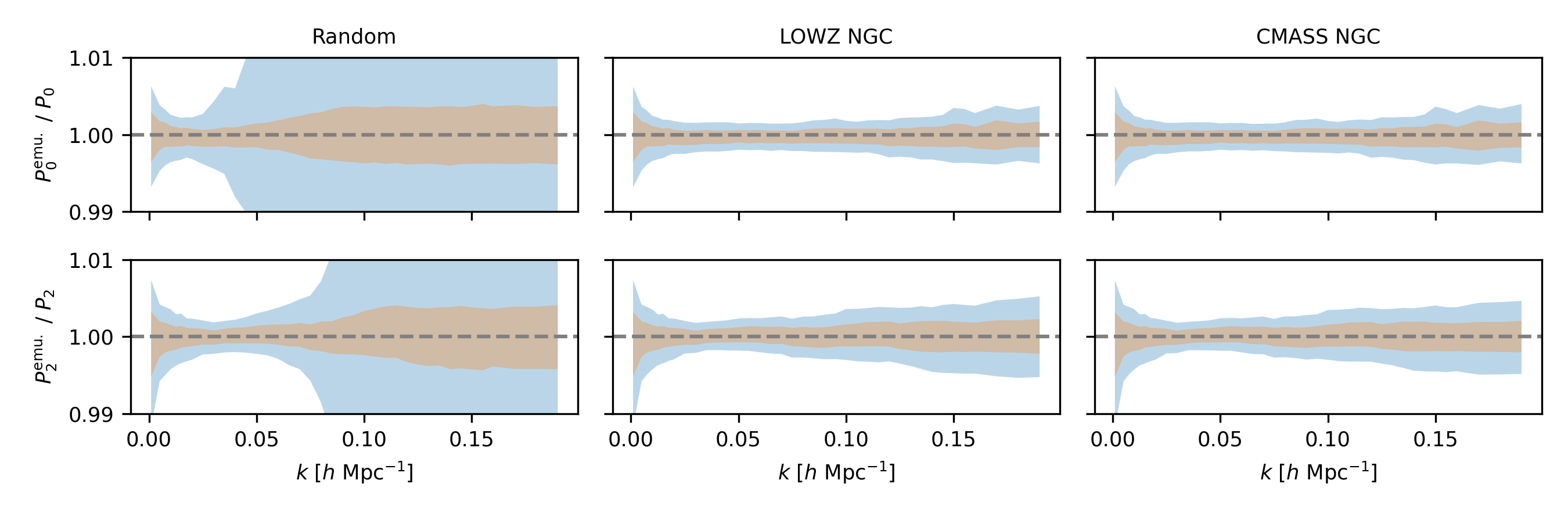}
    \caption{The prediction error at the power spectrum level of the \texttt{EFTEMU}. Each panel shows the ratio of the \texttt{EFTEMU} to the \texttt{PyBird} predictions for the same 2000 test cosmologies uniformly sampled across the $\Lambda$CDM parameter space shown in figure \ref{fig:sample_space}. In each column different sets of bias parameters are used. For the first column 2000 random sets of bias parameters are sampled from the prior defined in equation 3.18 of \citet{damico_cosmological_2020}. The second and third columns use the best\textcolor{teal}{-}fit LOWZ NGC and CMASS NGC parameters from \citet{damico_cosmological_2020} respectively. The orange and blue shaded regions show the $1\sigma$ and $2\sigma$ regions respectively. The $2\sigma$ region for the random bias parameters reaches a maximum of $\sim3\%$ for $P_0$ and $\sim5\%$ for $P_2$. This is the prediction accuracy for the \texttt{EFTEMU} trained at $z=0.38$; plots for the other two redshifts considered look very similar.}
    \label{fig:per_err}
\end{figure*}

We assess the prediction accuracy by examining the ratio of the \texttt{EFTEMU} predictions to the \texttt{PyBird} predictions. These are shown in figure \ref{fig:per_err}. The orange and blue shaded regions show the $1\sigma$ and $2\sigma$ regions respectively. We can see that at the $1\sigma$ level the \texttt{EFTEMU} is producing predictions with an error of $< 1\%$ for all sets of bias parameters. This is $>1\%$ for the random bias parameters for scales $k\gtrsim 0.05\ h\ \mathrm{Mpc}^{-1}$ for $P_0$ ($k\gtrsim 0.08\ h\ \mathrm{Mpc}^{-1}$ for $P_2$). However for the best-fit bias parameters the prediction error is still $<1\%$ at the $2\sigma$ level. These results are indicating that the \texttt{EFTEMU} is capable of producing predictions for $P_0$ and $P_2$ with better than 1\% accuracy for multipoles that are BOSS-like, although there are some regions of the prior space in which the prediction error is considerably higher. The $2\sigma$ region for the random bias parameters in figure \ref{fig:per_err} reaches a maximum of $\sim3\%$ for $P_0$ and $\sim5\%$ for $P_2$. We do note that in these regions of the bias prior space the multipoles look significantly different from those observed from BOSS data. The slightly higher prediction error for $P_2$ is a consequence of higher complexity of some of the $P_{n,2}$ components compared to the $P_{n,0}$ components; this higher complexity makes these components more difficult to predict. When bias parameters related to these components are drawn from the extremes of the bias prior (i.e. they are large) the prediction error of $P_2$ is higher.

\section{Mock Analyses}
\label{sec:mock_analyses}
In this section we describe a series of mock full shape analyses of the galaxy power spectrum multipoles. These analyses allow us to evaluate the performance of \texttt{EFTEMU} at the inference level rather than the power spectrum level. Conducting these analyses allows us to determine how the achieved level of prediction accuracy from \texttt{EFTEMU} impacts constraints on cosmological parameters.

\subsection{Mock power spectrum multipoles}
\label{subsec:mock_multipoles}

We produce a set of mock multipoles to facilitate the analyses of this section. These multipoles are produced using \texttt{PyBird}, and the Planck 2018 parameters as the true cosmology. We generate mock multipoles with the same cosmology at the three redshifts at which the \texttt{EFTEMU} has been trained. The bias parameters used to generate the mocks depend on the redshift. For the multipoles at $z=0.38$ we use the best\textcolor{teal}{-}fit LOWZ NGC parameters from \citep{damico_cosmological_2020}, whilst for the multipoles at $z=0.51$ and at $z=0.61$ we use the best-fit CMASS NGC parameters. We present the cosmological and bias parameters in table \ref{tab:mock_params}. It should be noted that the best-fit bias parameters are a result of analyses of BOSS LOWZ and CMASS samples with $z=0.32$ and $z=0.57$ respectively. There will be some redshift evolution of these bias parameters, however, we expect this redshift evolution to be small for $\Delta z \sim 0.05$ \citep{fry_evolution_1996, salazar-albornoz_clustering_2017}. Using these bias parameters will produce mock multipoles that are LOWZ-\textit{like} at $z=0.38$, and CMASS-\textit{like} at $0.51$ and $0.61$, but will not represent these two samples exactly. We produce mocks with 39 linearly spaced $k$-bins covering the range $0.001\ h\ \mathrm{Mpc}^{-1} \leq k \leq 0.19\ h\ \mathrm{Mpc}^{-1}$.\\

\begin{figure}
	\includegraphics[width=\columnwidth]{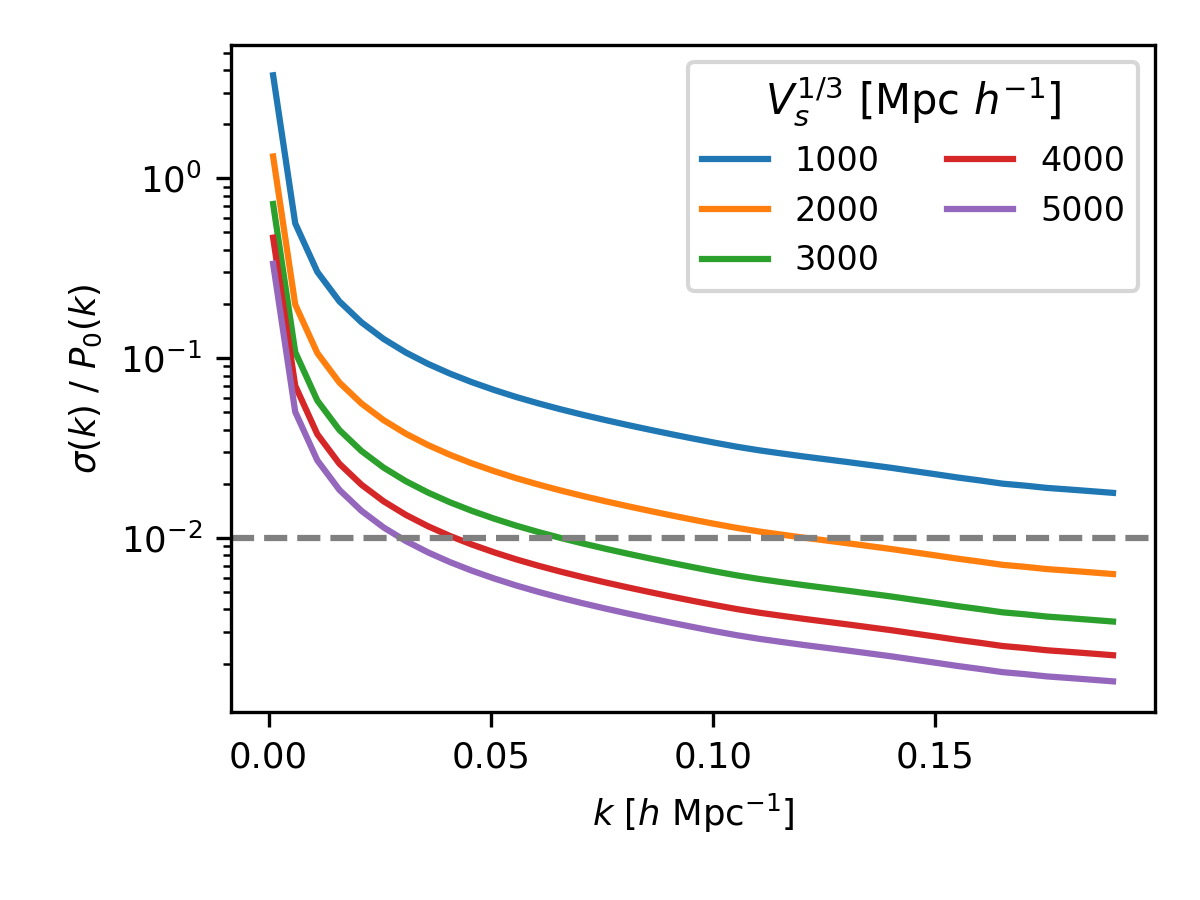}
    \caption{The ratio of the diagonal of the Gaussian covariance matrices described in section \ref{subsec:mock_multipoles} to the mock monopole at $z=0.38$. Each coloured line represents a covariance matrix with a different volume $V_s$. The grey dashed line shows the 1\% level. Plots at $z=0.51$ and $z=0.61$ look similar.}
    \label{fig:cov_per_err}
\end{figure}

We calculate Gaussian covariance matrices for these sets of mock multipoles using the equations presented in appendix C of \citet{taruya_baryon_2010}. We calculate several Gaussian covariance matrices for each set of mock multipoles, each with different mock survey volumes $V_s$. This allows us to investigate how the noise level of the mock multipoles  impacts the significance of any bias in the posterior predictions of this section. The values of $V_s^{1/3}$ considered are $[1000, 2000, 3000, 4000, 5000]\ \mathrm{Mpc} \ h^{-1}$. Figure \ref{fig:cov_per_err} shows the ratio of the diagonal of these covariances to the mock monopole at $z=0.61$. Each coloured line corresponds to a different $V_s$, whilst the grey line shows the 1\% level. We can see that for all but $V_s^{1/3} = 1000 h \ \mathrm{Mpc}^{-1}$ this ratio is below 1\% for some portion of the scales considered. The shot-noise in all covariance matrices comes from a $1/\bar{n}_g$ term. For the covariance at $z=0.38$ we use a value of $4.5\times10^{-4} \ (h \ \mathrm{Mpc}^{-1})$, and for the covariance at $z=0.51$ and $z=0.61$ we use a value of $4\times10^{-4} \ (h \ \mathrm{Mpc}^{-1})$. Figure \ref{fig:cov_per_err} looks similar for $z=0.51$ and $z=0.61$.

\begin{table*}
 \centering
 \begin{tabular}{c c c c c c c c c c c c c c c c} 
 \hline
 Redshift & $\omega_c$ & $\omega_b$ & $h$ & $\ln{\left(10^{10}A_s \right)}$ & $n_s$ & $b_1$ & $c_2$ & $b_3$ & $c_4$ & $c_{ct}$ & $c_{r,1}$ & $c_{r,2}$ & $c_{\epsilon,1}$ & $c_{\epsilon,\mathrm{mono.}}$ & $c_{\epsilon,\mathrm{quad.}}$ \\ 
 \hline
 0.38 & 0.11933 & 0.02242 & 0.6766 & 3.047 & 0.9665 & 1.73 & 1.0 & -1.0 & 0.0 & 0.2 & -10.03 & 0.0 & 0.0 & 0.0 & -2.1\\
 \hline
 0.51, 0.61 & 0.11933 & 0.02242 & 0.6766 & 3.047 & 0.9665 & 2.22 & 1.2 & 0.1 & 0.0 & 0.4 & -7.7 & 0.0 & 0.0 & 0.0 & -3.7\\
 \hline
 \end{tabular}
\caption{Cosmological and bias parameter values used to generate the mock multipoles described in section \ref{subsec:mock_multipoles}. The cosmological parameters are the Planck 2018 $\Lambda$CDM results, and the bias parameters are the best fit LOWZ NGC ($z=0.38$) and CMASS NGC ($z=[0.58,0.61]$) parameters from \citet{damico_cosmological_2020}. For detailed definitions of the bias parameters we refer the reader to sections 2.1 and 3.5 of \citet{damico_cosmological_2020}, and the references therein.}
\label{tab:mock_params}
\end{table*}

\subsection{MCMC with \texttt{EFTEMU}}
\label{subsec:MCMC_EFTEMU}

We calculate posterior distributions on cosmological and bias parameters via Markov Chain Monte Carlo (MCMC). We run all MCMCs with a \texttt{Python} implementation of ensemble slice sampling: \texttt{zeus} \citep*{karamanis_zeus_2021}. The \texttt{EFTEMU}, and all other emulators included with \texttt{matryoshka}, work very well with ensemble samplers like \texttt{zeus} as \texttt{TensorFlow} (the code with which all the emulators are built) has been heavily optimised for making batch predictions (i.e. making predictions for multiple sets of cosmological parameters).

We vary three out of the five cosmological parameters $\left[\omega_c, h, \ln{\left(10^{10}A_s\right)}\right]$, and seven out of the ten bias parameters $\left[b_1, c_2, b_3, c_{ct}, c_{r,1}, c_{\epsilon,1}, c_\mathrm{quad.}\right]$. We fix $n_s$ at its true value as we do not expect to get constraints much tighter than the prior. Rather than fixing $\omega_b$ to its true value we fix the baryon fraction $f_b = \omega_b/(\omega_c+\omega_b)$ to mimic the blind mock analyses presented in \citet{nishimichi_blinded_2020}. We fix $c_4$, $c_{r,2}$, and $c_\mathrm{mono.}$ all equal to 0, as in \citet{damico_cosmological_2020}. \citet{damico_cosmological_2020} notes that the components of the power spectrum prediction involving $c_4$ and $c_\mathrm{mono.}$ have a negligible impact on their analysis for a BOSS-like volume. They also note that as their analysis does not include the hexadecapole $P_4$ they can absorb the contribution of $c_{r,2}$ into $c_{r,1}$.

We use uniform priors on all cosmological parameters; they are presented in table \ref{tab:priors}. The width of these priors are determined by the training space used for the \texttt{EFTEMU}. It should however be noted that fixing $f_b$ has the effect of projecting the prior on $\omega_b$ onto the prior on $\omega_c$; we show this in table \ref{tab:priors} in parentheses. The priors on the bias parameters are the same as those in \citet{damico_cosmological_2020}, which are also presented in table \ref{tab:priors}. We use a Gaussian likelihood of the form
\begin{equation}
    \ln{\mathcal{L}(P|\theta,\phi)} = -\frac{1}{2}(P-\tilde{P})^T\bm{C}^{-1}(P-\tilde{P})\ ,
    \label{eq:likelihood}
\end{equation}
with $P$ being the mock data $P=[P_0, P_2]$, $\tilde{P}$ being the predictions $\tilde{P}=[\tilde{P}_0, \tilde{P}_2]$ for a given set of cosmological parameters $\theta$ and bias parameters $\phi$, and $\bm{C}$ being the covariance matrix.

\begin{table}
 \centering
 \begin{tabular}{c c} 
 \hline
 Parameter & Prior\\ 
 \hline
 $\omega_c$ & $\mathcal{U}(0.101, 0.140)$ \{$\mathcal{U}(0.110, 0.129)$\} \\
 $h$ & $\mathcal{U}(0.575, 0.748)$ \\
 $\ln{\left(10^{10}A_s\right)}$ & $\mathcal{U}(2.78, 3.32)$ \\
 \hline
 $b_1$ & $\mathcal{U}(0, 4)$ \\
 $c_2$ & $\mathcal{U}(-4, 4)$ \\
 $b_3$ & $\mathcal{N}(0, 2)$ \\
 $c_{ct}$ & $\mathcal{N}(0, 2)$ \\
 $c_{r,1}$ & $\mathcal{N}(0, 8)$ \\
 $c_{\epsilon,1}/\bar{n}_g$ & $\mathcal{N}(0, 400)$ \\
 $c_\mathrm{quad.}$ & $\mathcal{N}(0, 2)$ \\
 \hline
 \end{tabular}
\caption{Priors on the cosmological and bias parameters used in the full shape analyses described in section \ref{subsec:MCMC_EFTEMU}. $\mathcal{U}(a, b)$ denotes a uniform distribution with boundaries $a$ and $b$, and $\mathcal{N}(\mu, \sigma)$ denotes a normal distribution with mean $\mu$ and standard deviation $\sigma$. All other cosmological and bias parameters are fixed at their true values presented in table \ref{tab:mock_params}.}
\label{tab:priors}
\end{table}

Throughout this work we monitor the integrated autocorrelation time $\tau$ to judge convergence of the MCMCs. $\tau$ is an estimate of number of chain steps required to produce an independent sample from the posterior; see section 3 of \citet{foreman-mackey_emcee:_2013} for a definition of $\tau$ and argument for its use to measure sampling performance. We run chains for a minimum of 5,000 steps and then require that the length of the chains  are at least $100\tau$. We also examine the value of $\tau$ as the length of the chain increases. For a chain to be considered converged we require that $\tau$ has changed by <1\% in the previous 200 steps.

\subsubsection{Different redshifts}
\label{subsubsec:redshift}

The first set of analyses we run are designed to test the \texttt{EFTEMU} at each of the three redshift slices at which it has been trained. For these analyses we use the Gaussian covariance matrices with $V_s^{1/3}=5000 \ \mathrm{Mpc} \ h^{-1}$.

\begin{figure*}
	\includegraphics[width=\linewidth]{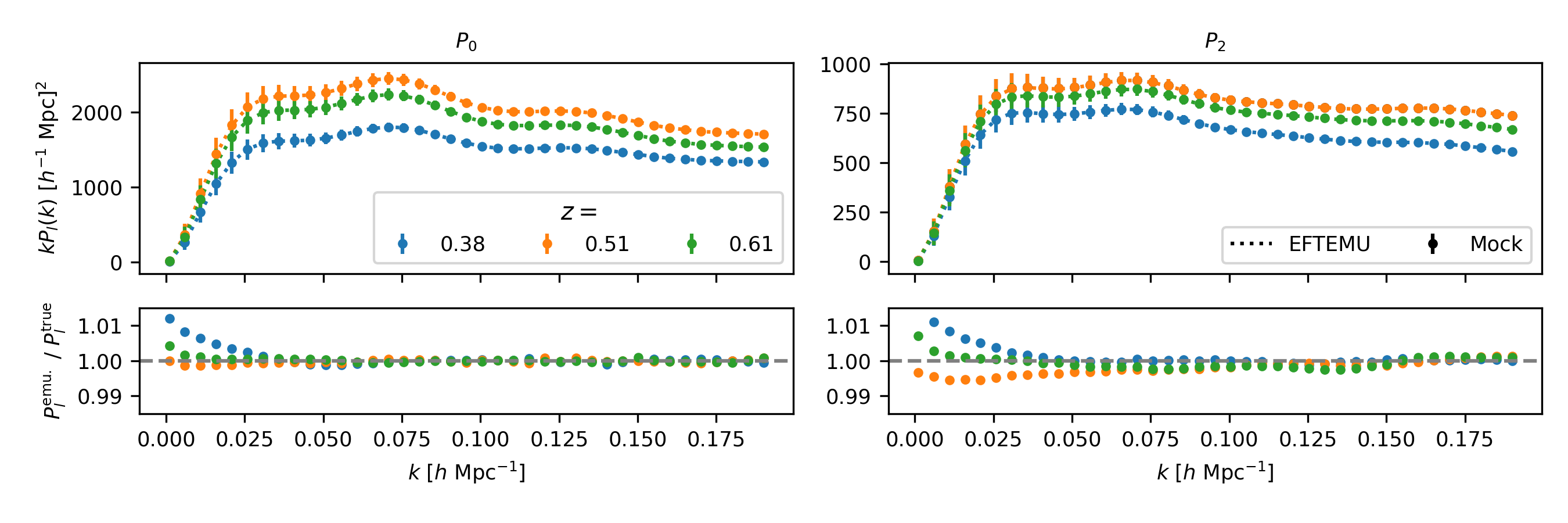}
    \caption{Comparison of the multipole predictions at the maximum \textit{a posteriori} (MAP) estimate to the mocks calculated with \texttt{PyBird} (see section \ref{subsec:mock_multipoles}). The MAP estimate was calculated with the \texttt{EFTEMU} and Gaussian covariance with $V_s=(5000\ \mathrm{Mpc} \ h^{-1})^3$. The coloured points and errorbars in the top panels show the mock data and the dotted lines show the \texttt{EFTEMU} predictions. The errorbars have been increased by a factor of five for $P_0$ to make them visible. The bottom panels show the ratio of the \texttt{EFTEMU} predictions to the mock data. The errorbars have been omitted for clarity.}
    \label{fig:map_predictions}
\end{figure*}

\begin{figure}
	\includegraphics[width=\columnwidth]{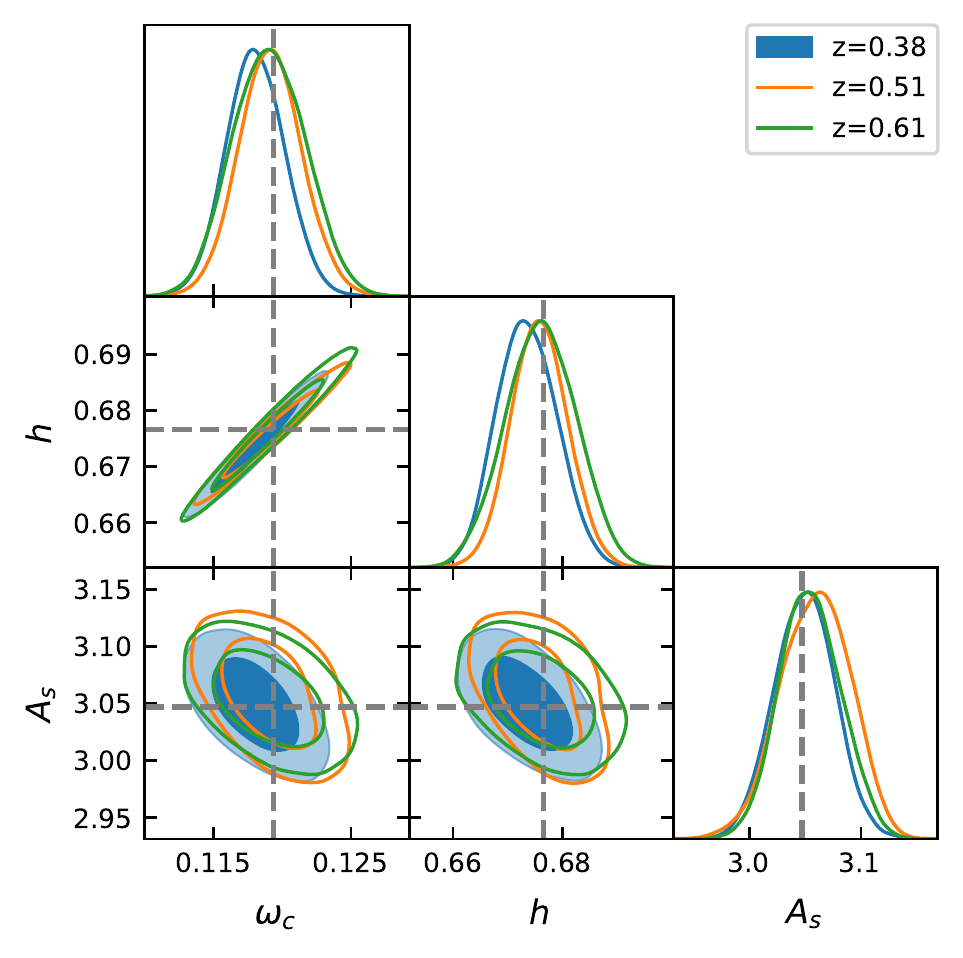}
    \caption{Corner plot showing marginalised 1D and 2D posterior distributions resulting from the full shape analyses described in section \ref{subsubsec:redshift}. The contour levels are $1\sigma$ and $2\sigma$. The grey dashed lines shows the location of the true cosmological parameters used to generate the mock data. A total of ten parameters are included in the MCMC; however the marginalised posteriors for the bias parameters are not shown as they depend on redshift.}
    \label{fig:corner_redshift}
\end{figure}

Figure \ref{fig:map_predictions} compares the \texttt{EFTEMU} predictions for the maximum \textit{a posteriori} (MAP) estimates at each redshift to the mock multipoles. The MAP estimates are found by simply minimising the negative of the log probability (the sum of the log prior and the log likelihood). Both panels show that the predictions from \texttt{EFTEMU} agree with the mock multipoles very well, the emulators have not introduced any unexpected features in the multipoles and any deviation is significantly lower than the $1\sigma$ level. The bottom panel shows the ratio of the predictions to the mock data. We can clearly see that the predictions agree with the mocks with a $\lesssim1\%$ error on all scales considered. Figure \ref{fig:corner_redshift} shows the posterior distributions for the cosmological parameters resulting from the MCMC analyses at each redshift\footnote{All corner plots showing posterior distributions in this work were produced with \texttt{GetDist} \citep{lewis_getdist_2019}. Unless otherwise stated the smoothing parameters used were \texttt{smooth\_scale\_2D=0.4} and \texttt{smooth\_scale\_1D=0.2}.}. The two contour levels are $1\sigma$ and $2\sigma$, and the true values are shown with the grey dashed lines. We can see that there is good agreement between the posterior distributions, and that the truth is recovered within $1\sigma$ in all cases. There are slight shifts in the posterior peaks when looking at the marginalised 1D posteriors. There are a few possible sources for these shifts: the first and most relevant for this work is prediction error in the multipoles coming from the emulator, the second being influence of the prior. Determining how much of the shift is due to the emulator error requires comparison to the posterior that would be calculated with \texttt{PyBird}. We present comparisons with \texttt{PyBird} posteriors in section \ref{subsubsec:pybird}.

\subsubsection{Different volumes}
\label{subsubsec:volumes}

The next set of analyses test the \texttt{EFTEMU} with increasing $V_s$ (and thus increasing signal-to-noise). We analyse the same set of mock multipoles, at $z=0.38$, using Gaussian covariance matrices calculated with $V_s$ ranging from $1000\ \mathrm{Mpc}\ h^{-1}\leq V_s \leq 5000\ \mathrm{Mpc}\ h^{-1}$.

\begin{figure}
	\includegraphics[width=\columnwidth]{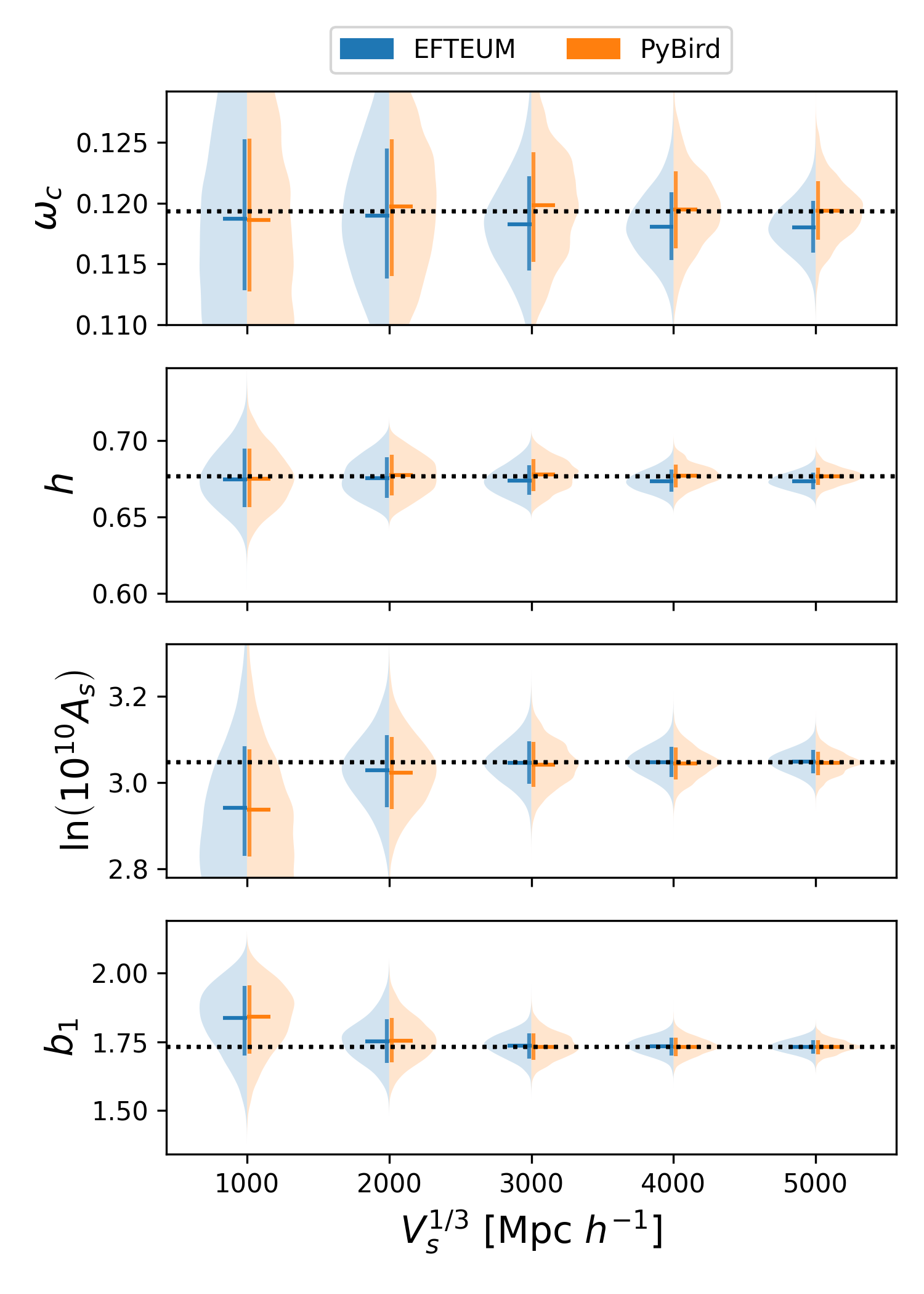}
    \caption{Violin plots showing the marginalised 1D posteriors on the cosmological parameters and linear bias $b_1$ resulting from the analyses of section \ref{subsubsec:volumes}. The horizontal coloured lines represent the median of the marginalised posterior, whilst the vertical coloured lines represent the $1\sigma$ region. Blue shows the posteriors calculated with the \texttt{EFTEMU}, whilst orange shows the corresponding \texttt{PyBird} posterior calculated via the method described in section \ref{subsubsec:pybird}. The dashed black lines show the truth values. At each volume the width of the shaded region represents the level of probability, narrower regions represent lower probability, whilst wider regions represent higher probability. For $\omega_c$ and $\ln{\left(10^{10}A_s\right)}$ the $y$-axis is limited to the width of the prior.}
    \label{fig:violins}
\end{figure}

Figure \ref{fig:violins} shows the marginalised 1D posteriors on the cosmological parameters and linear bias $b_1$ resulting from these analyses in blue. We can see that for small volumes there is no significant constraint beyond the prior on $\omega_c$. This is due to the projection of the $\omega_b$ prior onto $\omega_c$ mentioned in section \ref{subsec:MCMC_EFTEMU}. We can also see that there is a shift in the median value of the $\ln{\left(10^{10}A_s\right)}$ posterior for low $V_s$. This shift is prior volume projection effect that impacts the marginalised posteriors \citep*{carrilho_cosmology_2022, simon_consistency_2022}. Figure 3 of \citet{simon_consistency_2022} shows how an increase in $V_s$ can reduce the impact of these prior effects on the marginalised posteriors. The same can be seen by observing how the posteriors change with volume; as $V_s$ increases the constraints on all cosmological parameters tighten whilst remaining consistent with their true values. Small shifts can be seen in the $\omega_c$ and $h$ marginalised posteriors at all volumes. These shifts are due to emulator errors and not prior effects as they do not resolve with increasing $V_s$. We do however note that there is still consistency with the truth at the $1\sigma$ level for all $V_s$. These results are demonstrating that the prediction accuracy from the \texttt{EFTEMU} is sufficiently high so as not to induce any significant bias when performing inference on a sample with $V_s\leq 5000\,\mathrm{Mpc}\ h^{-1}$.

\subsubsection{Comparison with \texttt{PyBird}}
\label{subsubsec:pybird}

Finally we focus on a comparison of the \texttt{EFTEMU} and \texttt{PyBird}. The MCMCs of the previous sections required $\mathcal{O}(10^6)$ likelihood evaluations to reach convergence. Running a single MCMC with \texttt{PyBird} would require considerable resources. A more computationally efficient alternative method that is suited to the comparison we want to make here is \textit{importance sampling}. Importance sampling allows us to estimate a target distribution $P(\theta)$ by sampling from a proposal distribution $Q(\theta)$ and then applying an importance weight $I(\theta)$ to each of these samples. The weights $I(\theta)$ correspond to the ratio $P(\theta)\ / \ Q(\theta)$. Our target distribution is the \texttt{PyBird} posterior, and our proposal is the \texttt{EFTEMU} posterior. The MCMC chains calculated with the \texttt{EFTEMU} represent samples from $Q(\theta)$. Weights are calculated for a subset of the chain samples by computing the ratio of the \texttt{PyBird} likelihood to the \texttt{EFTEMU} likelihood for each sample\footnote{We want to consider the same prior for these comparisons as such the ratio of likelihoods is equivalent to the ratio of posterior probabilities.}.

We do not expect the peak of the \texttt{PyBird} posterior to be significantly shifted compared to that calculated with the \texttt{EFTEMU}; in the previous sections we have shown that the \texttt{EFTEMU} posteriors agree with the truth at the $1\sigma$ level. This recovery of the truth does not however indicate how the width of the distribution could have been impacted. In the scenario in which the proposal $Q(\theta)$ is narrower than the target $P(\theta)$ the calculation of the weights $I(\theta)$ can be numerically unstable. This can cause samples to be assigned unrealistically high weights and this can then distort the resulting \texttt{PyBird} posterior. To mitigate against this when calculating \texttt{PyBird} posteriors for $V_s^{1/3}\ge 3000\ \mathrm{Mpc}\ h^{-1}$ we draw samples from \texttt{EFTEMU} posteriors lower values of $V_s$.\footnote{For almost all cases in which we follow this procedure the proposal posterior is calculated with $V_s^\mathrm{prop.}=V_s^\mathrm{targ.}-1000\ \mathrm{Mpc}\ h^{-1}$. The only exception is the case where $z=0.38$ and $V_s^\mathrm{targ.}=5000\ \mathrm{Mpc}\ h^{-1}$; in this case $V_s^\mathrm{prop.}=3000\ \mathrm{Mpc}\ h^{-1}$.} The lower signal-to-noise ratio associated with lower volumes means these posteriors should be wider than the target \texttt{PyBird} posterior.

\begin{figure*}
	\includegraphics[width=\linewidth]{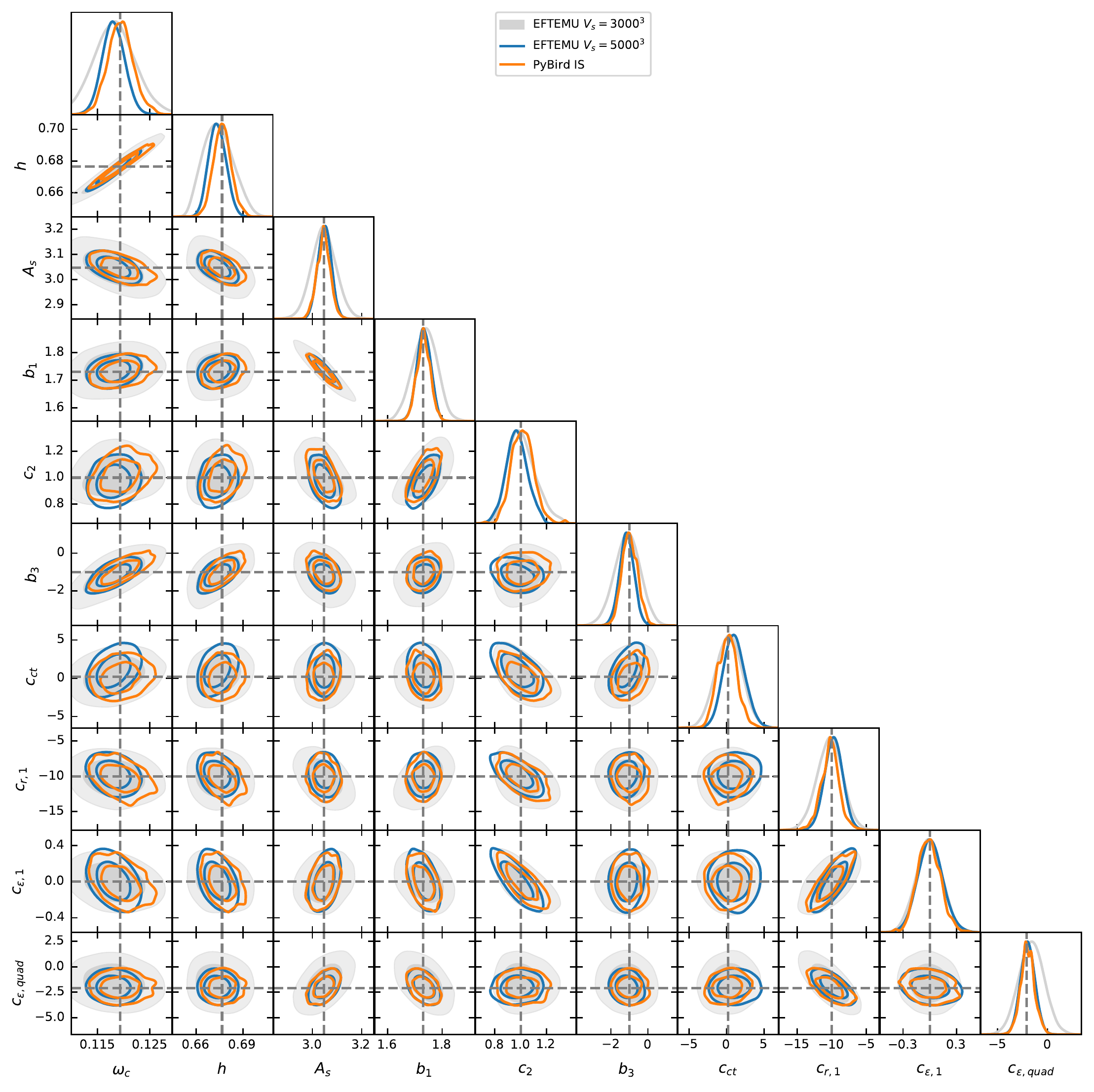}
    \caption{Corner plot showing marginalised 1D and 2D posterior distributions resulting from the full shape analyses described in section \ref{subsubsec:pybird}. The contour levels are $1\sigma$ and $2\sigma$. The grey contours show the proposal distribution used to calculate the \texttt{PyBird} posterior via importance sampling (IS). The grey dashed line shows the location of the true cosmological parameters used to generate the mock data. Both of the \texttt{EFTEMU} sets of contours are calculated via MCMC. The \texttt{PyBird} contours are obtained via importance sampling of the \texttt{EFTEMU} chains.}
    \label{fig:corner_pybird}
\end{figure*}

Figure \ref{fig:corner_pybird} shows a comparison of the \texttt{EFTEMU} and \texttt{PyBird} posteriors for all parameters varied in our MCMC analysis of the mock data at $z=0.38$ with $V_s^{1/3}=5000\ \mathrm{Mpc}\ h^{-1}$. We have also plotted the \texttt{EFTEMU} posterior that was used as the proposal distribution when calculating the \texttt{PyBird} posterior. The first thing to note is the significant difference between proposal distribution and the \texttt{PyBird} contours. In many of the 2D projections we can clearly see that the $1\sigma$ region of the proposal encompasses the $2\sigma$ region of the \texttt{PyBird} posterior. The second thing to note is that the agreement between the \texttt{EFTEMU} and \texttt{PyBird} posteriors is very good, and they are almost indistinguishable in many of the 1D and 2D projections. There are however some clear differences; the peaks of the marginalised 1D posteriors for $\omega_c$ and $h$ are shifted from the truth, whilst the \texttt{PyBird} posterior peaks agree with the truth as expected. These differences can also be seen in figure \ref{fig:violins}. The difference in the median values of the marginalised posteriors is $\lesssim0.6\sigma$, whilst the width of the 68\% credible intervals of the posteriors agree within $\sim10\%$. The only source for these differences is prediction errors from the emulator as all other aspects of the analysis pipeline are the same. It should however be noted that these results represent the "worst case scenario". At lower volumes, where the signal-to-noise ratio of the mock is lower and the emulator errors are less consequential, these differences are less significant. When analysing the mocks at higher redshift these differences are also less significant. A table quantifying the difference between the \texttt{EFTEMU} and \texttt{PyBird} posteriors for each full shape analysis of this work is presented in appendix \ref{sec:tab_comp}. These results indicate that any prediction errors from the \texttt{EFTEMU} are not inducing any significant bias in the posteriors on cosmological and bias parameters.

\section{Computational Performance}
\label{sec:performance}
The prediction speed of the \texttt{EFTEMU} will be processor-dependent. Any computational metrics discussed in this section and throughout the paper refer to the scenario in which predictions are being made on a laptop with an Intel i5 2.50 GHz dual-core processor with four threads and 8 GB of RAM.

To assess the prediction speed of the \texttt{EFTEMU} we make predictions for $N$ sets of cosmological and bias parameters 100 times and report the mean and standard deviation. In the case where $N=1$ the prediction speed is $4.22\ \mathrm{ms} \pm 474\ \mathrm{\mu s}$ per prediction per multipole. However when making predictions on a batch where $N=20$ ($N=100$) we find a prediction speed of $0.280\ \mathrm{ms} \pm 46.1\ \mathrm{\mu s}$ ($0.104\ \mathrm{ms} \pm 8.12\ \mathrm{\mu s}$) per prediction per multipole. This highlights the increased efficiency of the \texttt{EFTEMU} when making batch predictions and we reiterate that this property is very beneficial when using ensemble samplers such as \texttt{zeus}. If we repeat this test for \texttt{PyBird} we find a prediction speed of $1.01\ \mathrm{s} \pm 13.3\ \mathrm{ms}$ per prediction per multipole. Thus using the \texttt{EFTEMU} results in a speed up by a factor of $240 \pm 27.1$ when making single predictions, and a factor of $3620 \pm 597$ ($9730 \pm 770$) when making predictions on a batch of input parameter sets.

\section{Conclusions}
\label{sec:conclusions}
In this work we expanded the \texttt{matryoshka} suite of neural-network-based emulators with the \texttt{EFTEMU}. We have trained the \texttt{EFTEMU} to predict bias independent components of the EFTofLSS model for the galaxy power spectrum multipoles. The \texttt{EFTEMU} achieves better than 1\% prediction accuracy ($2\sigma$ level) for both the monopole and the quadrupole on scales $0.001\ h\ \mathrm{Mpc}^{-1} \leq k \leq 0.19\ h\ \mathrm{Mpc}^{-1}$ for BOSS-like bias parameters, whilst producing very fast predictions. The \texttt{EFTEMU} can produce 10,000 predictions in $\sim 1.5\ \mathrm{s}$, for comparison this would take $\sim5.6\ \mathrm{hr}$ with \texttt{PyBird}. The \texttt{EFTEMU} is implemented in \texttt{Python} and is publicly available \url{https://github.com/JDonaldM/Matryoshka}.

We calculate sets of mock galaxy power spectrum multipoles with BOSS-like bias parameters, at $z=[0.38, 0.51, 0.61]$. Using these sets of multipoles we run a series of mock analyses designed to assess any potential bias when performing cosmological inference with the \texttt{EFTEMU}. We have shown that at each redshift the multipoles predicted by the \texttt{EFTEMU} at the MAP estimate agree with the mocks at better than 1\% for a mock volume of $V_s^{1/3}=5000\ \mathrm{Mpc}\ h^{-1}$. We have also shown that the true cosmology is recovered within $1\sigma$ at each redshift. As a further test of the \texttt{EFTEMU} we run mock analyses with $1000\ \mathrm{Mpc}\ h^{-1}\leq V_s \leq 5000\ \mathrm{Mpc}\ h^{-1}$, and verify that there is no significant bias introduced in the inferred cosmology with mock volumes up to $5000\ \mathrm{Mpc}\ h^{-1}$. As a final test we compare posterior distributions calculated with \texttt{EFTEMU} to those calculated with \texttt{PyBird}. We obtain the \texttt{PyBird} posterior by importance sampling the chains obtained with \texttt{EFTEMU} and verify that there is no significant bias when comparing the \texttt{PyBird} posterior and the \texttt{EFTEMU} posterior. We note that the method used for this comparison highlights potential synergies between emulators and the target models they are developed to approximate. Estimating the \texttt{PyBird} posterior with importance sampling on average required $\mathcal{O}(10^4)$ model evaluations, which is two orders of magnitude less than what would be required to run a typical MCMC to convergence. As an example, if the signal-to-noise ratio of a given measurement is higher than that which the emulator has been tested, the emulator could be used to rapidly generate a posterior with artificially inflated uncertainty on the measurement. This posterior can then be importance sampled with \texttt{PyBird} and the true measurement uncertainty to obtain the posterior orders of magnitude faster than running an MCMC with \texttt{PyBird} alone.

\section*{Acknowledgements}
For the purpose of open access, the author(s) has applied a Creative Commons Attribution (CC BY) licence to any Author Accepted Manuscript version arising. The authors would like to thank Minas Karamanis for the useful discussions regarding importance sampling and the \texttt{PyBird} developers for making their code public.
JD-M was supported by a STFC studentship.
KK is supported the UK STFC grant ST/S000550/1. FB has received funding from the European Research Council (ERC) under the European Union's Horizon 2020 research and innovation programme (grant agreement 853291, ``FutureLSS"). FB is a Royal Society University Research Fellow.

\section*{Data Availability}
The \texttt{EFTEMU} is publicly available as part of \texttt{matryoshka}, which can be found here \url{https://github.com/JDonaldM/Matryoshka}. We make available all training and test data generated for this work in a repository that can be found here \url{https://github.com/JDonaldM/matryoshka_II_paper}. We also included all mock multipoles and Gaussian covariance matrices, and Python scripts and notebooks that allow each step of the analysis presented here to be reproduced. 

\bibliographystyle{mnras}
\bibliography{ref}

\appendix
\section{Tabular comparison with \texttt{PyBird}}
\label{sec:tab_comp}
Table \ref{tab:tab_comp} quantitatively compares the \texttt{EFTEMU} and \texttt{PyBird} posteriors for the cosmological parameters of interest for all the mock analyses of this work. We compare the distributions by examining the normalised residual
\begin{equation}
    \Delta \theta = \frac{\theta^\texttt{EFTEMU}_{50}-\theta^\texttt{PyBird}_{50}}{0.5(\theta^\texttt{PyBird}_{84}-\theta^\texttt{PyBird}_{16})}\ ,
    \label{eq:median_diff}
\end{equation}
where $\theta^\texttt{EFTEMU}_{50}$ and $\theta^\texttt{PyBird}_{50}$ are the posterior median values for a given parameter $\theta$ measured from the \texttt{EFTEMU} and \texttt{PyBird} posteriors respectively. $\theta^\texttt{PyBird}_{84}$ and $\theta^\texttt{PyBird}_{16}$ are the 84th and 16th percentiles. We also examine the ratio of widths of the 68\% credible intervals
\begin{equation}
    \Sigma \theta = \frac{\theta^\texttt{EFTEMU}_{84}-\theta^\texttt{EFTEMU}_{16}}{\theta^\texttt{PyBird}_{84}-\theta^\texttt{PyBird}_{16}}\ ,
    \label{eq:width_ratio}
\end{equation} 
where $\theta^\texttt{PyBird}_{84}$ and $\theta^\texttt{PyBird}_{16}$ are the same as in equation \ref{eq:median_diff}, and $\theta^\texttt{EFTEMU}_{84}$ and $\theta^\texttt{EFTEMU}_{16}$ are the equivalent 84th and 16th percentiles calculated from the \texttt{EFTEMU} posterior.

\begin{table*}
 \centering
 \begin{tabular}{c c c c c c c c} 
 \hline
 Redshift & $V_s^{1/3}\ [\mathrm{Mpc}\ h^{-1}]$ & $\Delta \omega_c$ & $\Sigma \omega_c$ & $\Delta h$ & $\Sigma h$ & $\Delta \ln{\left(10^{10}A_s \right)}$ & $\Sigma \ln{\left(10^{10}A_s \right)}$ \\ 
 \hline
 0.38 & 1000 & 0.0169 & 0.988 & -0.0207 & 0.997 & 0.0314 & 1.03\\
 \hline
 0.38 & 2000 & -0.137 & 0.953 & -0.140 & 0.989 & 0.0636 & 0.995\\
 \hline
 0.38 & 3000 & -0.349 & 0.856 & -0.379 & 0.927 & 0.0806 & 0.946\\
 \hline
 0.38 & 4000 & -0.449 & 0.891 & -0.462 & 0.980 & 0.0819 & 0.960\\
 \hline
 0.38 & 5000 & -0.558 & 0.888 & -0.584 & 0.991 & 0.133 & 0.998\\
 \hline
 \hline
 0.51 & 1000 & -0.0117 & 0.992 & -0.00281 & 0.983 & 0.00501 & 1.01\\
 \hline
 0.51 & 2000 & -0.00213 & 0.984 & -0.0161 & 0.959 & 0.0587 & 1.03\\
 \hline
 0.51 & 3000 & 0.0316 & 0.964 & -0.0109 & 0.934 & 0.135 & 1.02\\
 \hline
 0.51 & 4000 & 0.0184 & 0.922 & -0.0332 & 0.911 & 0.156 & 1.05\\
 \hline
 0.51 & 5000 & 0.0662 & 0.954 & -0.024 & 0.947 & 0.103 & 1.13\\
 \hline
 \hline
 0.61 & 1000 & 0.0142 & 0.994 & 0.0191 & 1.00 & 0.0360 & 0.986\\
 \hline
 0.61 & 2000 & 0.0475 & 1.01 & 0.0608 & 1.05 & -0.0236 & 0.974\\
 \hline
 0.61 & 3000 & 0.0265 & 1.06 & 0.0540 & 1.11 & 0.0207 & 0.969\\
 \hline
 0.61 & 4000 & 0.0432 & 1.06 & 0.0496 & 1.14 & -0.0509 & 0.980\\
 \hline
 0.61 & 5000 & -0.0231 & 1.13 & -0.0459 & 1.20 & -0.0993 & 0.988\\
 \hline
 \end{tabular}
\caption{Quantitative comparison of the \texttt{EFTEMU} and \texttt{PyBird} posteriors. The $\Delta X$ columns show the normalised residual of the posterior median values for the cosmological parameters calculated via equation \ref{eq:median_diff}. The $\Sigma X$ columns show the ratio of the 68\% credible interval calculated via equation \ref{eq:width_ratio}.}
\label{tab:tab_comp}
\end{table*}


\bsp	
\label{lastpage}
\end{document}